**<u>TITLE</u>**

**Folding is Not Required for Bilayer Insertion: Replica Exchange Simulations of an α-Helical Peptide with an Explicit Lipid Bilayer**

**<u>AUTHORS</u>**

**Hugh Nymeyer**

Theoretical Biology & Biophysics Group, T-10
MS K710, T-10
Los Alamos National Laboratory
Los Alamos, NM 87545

PHONE:      505-660-5171
FAX:            505-665-3943
EMAIL:       hugh@lanl.gov

**Thomas B. Woolf**

Depts of Physiology and of Biophysics
Johns Hopkins University, School of Medicine
725 N. Wolfe St.
Baltimore, MD 21205

PHONE:      410-614-2643
FAX:            410-614-4436
EMAIL:       woolf@groucho.med.jhmi.edu

**Angel E. Garcia *(Corresponding Author)***

Theoretical Biology & Biophysics Group, T-10
MS K710, T-10
Los Alamos National Laboratory
Los Alamos, NM 87545

PHONE:      505-665-5341
FAX:            505-665-3943
EMAIL:       axg@lanl.gov

## SHORT TITLE

**Folding is not Required for Bilayer Insertion**

## KEYWORDS

# AB STRACT

We implement the replica exchange molecular dynamics algorithm to study the interactions of a model peptide (WALP-16) with an explicitly represented DPPC membrane bilayer. We observe the spontaneous, unbiased insertion of WALP-16 into the DPPC bilayer and its folding into an $\alpha$-helix with a trans-bilayer orientation. We observe that the insertion of the peptide into the DPPC bilayer precedes secondary structure formation. Although the peptide has some propensity to form a partially helical structure in the interfacial region of the DPPC/water system, this state is not a productive intermediate but rather an off-pathway trap for WALP-16 insertion. Equilibrium simulations show that the observed insertion/folding pathway mirrors the potential of mean force (PMF). Calculation of the enthalpic and entropic contributions to this PMF show that the surface bound conformation of WALP-16 is significantly lower in energy than other conformations, and that the insertion of WALP-16 into the bilayer without regular secondary structure is enthalpically unfavorable by 5-10 kcal/mol/residue. The observed insertion/folding pathway disagrees with the dominant conceptual model[1-3], which is that a surface bound helix is an obligatory intermediate for the insertion of $\alpha$-helical peptides into lipid bilayers. In our simulations, the observed insertion/folding pathway is favored because of a large (> 100 kcal/mol) increase in system entropy that occurs when the unstructured WALP-16 peptide enters the lipid bilayer interior. The insertion/folding pathway that is lowest in free energy depends sensitively on the near cancellation of large enthalpic and entropic terms. This suggests that intrinsic membrane peptides may have a diversity of insertion/folding behaviors depending on the exact system of peptide and lipid under consideration.

Membranes and membrane proteins are dynamically active systems involved in essential biological processes. Whole genome analysis indicates that 20-30% of all open reading frames code for membrane spanning α-helical bundle proteins[4,5]. Proteins with β-barrel architectures (e.g. porins) are coded for by several percent of the open reading frames in bacteria[6] and an unknown fraction in eukaryotic organisms. Many other proteins involved in cell-cell adhesion, immune recognition, and signal transduction also have single α-helical membrane spanning domains[7]. Because of difficulties in isolating, purifying, and crystallizing membrane proteins, only about 82 unique intrinsic membrane protein structures are known[3,8] at atomic resolution compared with the thousands of globular proteins that have been solved[9]. Consequently, the protein-protein and protein-lipid interactions that stabilize intrinsic membrane proteins are not as well understood as the interactions that stabilize globular proteins. Prediction of membrane protein structure, of membrane protein folding, and of membrane protein dynamics is limited by our understanding of these protein-lipid interactions and lipid dynamics[10].

Because of these difficulties, model systems have been instrumental for understanding the general principles governing membrane protein structure and dynamics. An important model system has been the WALP series of peptides, which have an alternating, variable length alanine/leucine core flanked on both termini by two tryptophan residues[11-13]. These peptides have been demonstrated to form trans-membrane α-helices by CD[11], NMR[11,14,15], UV-Vis spectroscopy[14], transmission and atomic force microscopy[16,17], and mass spectrometry[18-20]. The compensatory changes in lipid structure induced by WALP peptides have also been studied via NM[11,20-26], electron spin resonance[21,22], microscopy[24], x-ray diffraction[27], and calorimetry[25]. These experimental studies have been complemented by molecular dynamics calculations, which have attempted to discern what lipid and peptide structural adjustments might occur for different length WALPs and different bilayer settings[28].

In this paper we report all-atom simulations of the interactions of a 16 residue WALP peptide with a solvated DPPC bilayer. Our simulations show the spontaneous insertion and folding of this WALP-16 peptide into the DPPC bilayer. These simulations are the first to show the unbiased, spontaneous insertion and folding of a hydrophobic peptide into an explicitly represented lipid bilayer. The spontaneous insertion and folding of peptides into trans-bilayer configurations is difficult to observe, because most membrane spanning peptides are highly hydrophobic and thus prone to aggregation. To our knowledge only three experimental kinetic studies of spontaneous peptide insertion processes exist[29-31]. Although no kinetic measurements of WALP insertion have been made, generic models of peptide insertion and folding have been constructed based on thermodynamic arguments[1-3]. Our simulated insertion does not agree with these models, suggesting that other previously discounted thermodynamic effects in the lipid component may alter the insertion process in some peptide/bilayer systems.

Our simulations have been conducted using 1024 processors on the Q-machine, a parallel computer at Los Alamos National Laboratory which at the time of this simulation is ranked as the third fastest in the world[32]. We have implemented a replica exchange (parallel tempering) molecular dynamics algorithm[8] on this machine. Replica exchange algorithms[33-37] were developed to study glassy systems with long relaxation times and are widely used in the context of protein folding[38-40] (reviewed by Nymeyer et al.[41]). In these methods, multiple copies or replicas of the same system are simulated in parallel at different temperatures, and temperatures are periodically exchanged between two replicas in a manner that preserves detailed balance. These algorithms speed equilibration by a large factor (perhaps 100x or more)[42-44] and enable us to observe insertion of a WALP peptide while simultaneously computing the equilibrium properties of the WALP/DPPC bilayer system.

As described in the methods section, we run three simulations. The first simulation begins with the peptide in a water solvated conformation. This simulation shows WALP spontaneously moving into conformations in which it is anchored into the bilayer. The second simulation starts with the peptide in an anchored conformation. This simulation shows four separate events in which WALP spontaneously inserts completely into the bilayer and forms α-helical secondary structure. The third simulation, which begins with the peptide inserted completely in the bilayer, is used to generate an equilibrium ensemble of WALP/DPPC conformations and measure the changes in thermodynamic quantities as a function of peptide structure and location in the bilayer.

The first two simulations suggest an insertion mechanism for WALP with three steps. In the first step, the peptides move into a membrane-anchored conformation. In this conformation, the peptide has inserted one or more of its Trp residues into the bilayer below the phosphate groups. These Trp residues anchor the peptide to the bilayer, although the peptide itself is still mostly water solvated. In the second step, the peptides insert into the lipid bilayer in an unstructured state, occupying the volume exposed in the lipid due to fluctuations of the lipid chains. In the third step these peptides form a helical nucleus, from which the whole α-helix rapidly forms. This helix orients roughly normal to the bilayer surface. These basic steps are observed in all the WALP peptides that inserted and folded. No WALP's were observed to insert directly from a surface bound helical conformation.

One of the observed insertion trajectories is shown in figure 1. Steps in the trajectory are shown along with a projection of the trajectory onto the plane spanned by distance of the WALP from the central bilayer plane and helical content of the WALP peptide. This trajectory is superimposed upon the potential of mean force (PMF). The trajectory of the inserted peptides mirrors the underlying potential of mean force determined in equilibrium.

The PMF at this temperature has three dominant basins of attraction. In the first basin, the peptide is mostly water solvated and non-helical but possibly anchored via one or more Trp residues into the membrane. In the second basin, the peptide is located in the bilayer interface and non-helical. This basin is a mixture of states in which the peptide is lying approximately in the plane of the interface and states in which the peptide is approximately normal to the interface. In the third basin, the peptide is helical and inserted in the membrane. The principal barrier to folding occurs during nucleation of the peptide in the center of the lipid bilayer. Although our PMF is shown for temperatures greater than the experimental conditions, extrapolation to lower temperatures does not appear to alter the insertion mechanism; however, the barrier to insertion does increase with decreasing temperature.

By fitting the temperature variation of our PMF, we have estimated the enthalpic and entropic changes of our system with peptide structure and position. These results (Figure 2) agree with the known thermodynamics of peptides interacting with lipid bilayers. In particular, we find that the insertion of the WALP-16 peptide into the DPPC bilayer in a largely unstructured state entails a significant enthalpic penalty of between 5-10 kcal/mol per residue. This is in agreement with calculations[45] and measurements on model compounds[46,47] that provide estimates of an enthalpic backbone desolvation penalty of 6-8 kcal/mol per residue. As expected, the enthalpy decreases sharply by nearly the same amount with the growth of hydrogen bonds along the α-helix. Although helix formation in water is generally enthalpically favored, helix formation followed by insertion may be less enthalpically favorable or even unfavorable, since hydrogen bonds in a fully formed α-helix retain some residual electrostatic interaction with the surrounding solvent[48].

Waters bound to the WALP may play an important role in stabilizing the peptide when inserted in the bilayer. Although we have prevented waters from penetrating to the center of the bilayer via a weak mean-field potential (see Methods for details), we still observe a significant amount of water bound to the WALP backbone prior to α-helix formation (see figure 4). Bound waters are certain to make insertion of the unfolded peptide more favorable than would be expected based on complete backbone desolvation. Figure 2 indicates that initially, insertion of the WALP is not strongly disfavored by enthalpy, presumably due to the presence of these bound waters.

From figure 2 we see that the surface bound partly helical conformations of the peptide are exceptionally low in energy. There are few experimental measurements of enthalpy changes upon the binding of small molecules to lipid bilayers. Jacobs and White[2] found negligible enthalpy changes upon binding of the small peptides Ala-X-Ala-O-*tert*-butyl (X = Leu, Phe, Trp) to DMPC vesicles. Similar results were found the COX IV peptide[49] and for several Trp derivatives[50]. In contrast, many aromatic amphiphiles have negative enthalpy changes upon binding to lipid bilayers[3,51-53]. DeVido et al.[54] suggest that negative enthalpic changes are generic for spontaneous binding of small molecules to ordered lipid chain phases. Figure 5 indicates that a strong Coulombic interaction can exist between the TRP residues which cap WALP and the phosphatidylcholine groups. This strong interaction is consistent with the experiments and statistical studies[55] that suggest favorable enthalpic interactions between Trp side chains and the DPPC/water interface. It is also expected that $\alpha$-helical hydrogen bonds will be more enthalpically favorable in the interface, because its effective dielectric constant is reduced from that of bulk water (probably to an $\varepsilon$ of $\approx$18) [3,56,57].

Although the partially folded helical surface bound conformations of WALP are low in energy, no passage directly these conformations to a trans-membrane helix is observed, suggesting that the surface bound state is acting more as an off-pathway trap than an intermediate for WALP insertion. Peptides other than WALP may be more stabilized as interfacial helices. Stronger stabilization of interfacial helices should favor insertion directly from a surface bound helical conformation.

The observed insertion behavior and the equilibrium potential of mean force disagree with the accepted hypothesis about the spontaneous insertion of peptides into lipid bilayers[1-3]. The accepted hypothesis, known as the four stage model (figure 3), posits that insertion into the membrane interior should occur only after significant secondary structure has already formed. This conclusion is based on calculations[45] and measurements[46-47] using bulk hydrophobic solvents, which show that the insertion of an unstructured peptide into the lipid interior will entail a high enthalpic cost due to the desolvation of the peptide backbone. This desolvation penalty is mostly absent in regularly structured peptides, because the backbone is already desolvated to a large extent, and the peptide hydrogen bond donors and acceptors are satisfied internally.

Our simulations are in agreement with the calculations and measurements suggesting a large penalty for backbone desolvation. However, our simulations also show that entropic changes are strong enough to overcome this desolvation penalty. No reliable estimates or measurements of the entropic changes due to the insertion of an unstructured peptide into a lipid bilayer exist to which we may compare our simulations; however, the observed entropic compensation effect is too large to be purely a simulation artifact.

Our molecular dynamics simulation results depend significantly on the force field used for the protein and the lipid and on the simulation conditions such as constant volume, constant cross sectional surface area, and system size. For example, we observe a transition to an ordered, gel like tilted phase for the lipid below 400K, while the transition should occur at 314 K. However, our simulations provide a molecular view of the folding of a transmembrane α-helix within an explicit lipid bilayer and suggest that the insertion and folding of peptides into lipid bilayers might be more complex than suggested. Simulations of peptides that are known to co-exist in the water and lipid phases[58,59] are under way. We expect our simulations to motivate experimentation of the time resolved kinetics of the insertion and folding of peptides in membranes.

In conclusion, our simulations provide a molecular view of the folding of a transmembrane α-helix within an explicit lipid bilayer. Our results show that the folding process might be more complex and subtle than suggested by the four stage model. In particular, we observe large entropic changes in the system that may make insertion of peptides into the bilayer prior to secondary structure formation favorable. The composition of the peptide and lipid are certain to modulate the large enthalpic and entropic terms driving insertion, making insertion mechanisms more variable than have been suggested by the four stage model. In this regard we suggest that membrane proteins, like globular proteins, may have multiple folding routes best described as motion on a multi-dimensional free energy surface[60,61]. More experimental studies of the folding pathways for α-helical monomers and dimers would be useful to further understand the molecular interplay that occur within the heterogenous solvation setting of the membrane bilayer and its associated waters.

## Methods

### Initial conditions

The starting point of our simulations is a fully solvated WALP-16 peptide [CHO-ALA-TRP$_2$-(LEU-ALA)$_5$-TRP$_2$-ALA-Ethanolamine] with 1048 TIP3P waters and a bilayer with 18 DPPC molecules in each monolayer. The initial conditions of the lipid were derived from previous simulations of the WALP-16 in a DPPC bilayer[8]. The WALP-16 peptide was placed initially into an unfolded water-solvated conformation with its long axis approximately horizontal to the membrane plane. The surface are per lipid in this initial conformation is 68 A$^2$. The surface are per lipid in pure DPPC bilayers has been measured[62] to be 64 A$^2$; however, the surface area for a mixed DPPC/WALP system is not accurately known. We assumed a constant surface area throughout the insertion process. This surface area was chosen to be match the result of another DPPC/WALP simulation performed with the same force-field[5].

**Simulations**

Simulations were carried out via a modified version of the CHARMM (version 28) program[63]. The force-field/energy function was the CHARMM22 all-atom force-field of Schlenkrich et al.[64,65].

The initial simulation was a 1ns replica exchange simulation using 38 replicas of the system with temperatures exponentially spaced from 350-505.8K. This was followed by a 1.6ns simulation involving 64 replicas exponentially spaced from 350K to 800K, where the simulation continued from a membrane anchored conformation obtained at the end of the initial 1ns simulation. Our lowest temperature is above the experimentally known phase transition of the DPPC bilayer (~314K).

For N replicas, exchanges are attempted between $N^2$ randomly chosen pairs of replicas at intervals of 250 integration steps. Integration steps are 2fs for the first 1.4 ns; 0.8fs for the next 200ps; and 1fs for the remainder. All bonds involving water are fixed for the first 1.6ns; all bonds involving hydrogen are fixed for the last 1ns. PME is used for the electrostatics with a 32x32x64 grid with fourth order spline interpolation, a 4A width Gaussian screening charge, and a 10-12A switching function on the direct interaction. The ensemble is NVT in a 35x35x72.75A box with a Nose-Hoover thermostat with a mass of 500 a.u..

A planar restraint is placed on the average position of the $C_2$ atom in each DPPC monolayer to maintain the membrane structure at high temperature. The minima are at +18.3555A and −18.2203A. The restraining potential is zero within 0.25A of the minimum and quadratic with a 10kcal/mol/A$^2$ force constant outside this region. A quadratic restraint of the form 0.2kcal/mol-A$^2$ x D$^2$ x (D$^2$-2.25A$^2$) applies to the water oxygens where D=z-25A for z>0 and D=z+25A for z<0. The effect of these constraints on the equilibrium properties of the lipid appears to be minimal since the calculated surface area tension of 35 dyn/cm at low T is consistent with the values reported for the force field used in our calculations[66]. The calculated surface tension decreases as T increases and becomes negative at T> 430 K.

The equilibrium simulation has identical settings. All replicas are started inserted but are non-helical. Step size is 0.8fs for the first 200ps with only the waters held rigid and 1fs for the remainder with all bonds involving hydrogen held fixed. Total simulation time is 3.5ns per replica. The final 1.5ns is used for analysis.

**Analysis**

The potential of mean force was computed for each temperature as −RTln(N), where N is the number of counts per bin at temperature T in Kelvin, and R is the ideal gas constant in units of kcal/mol/K. The ordinate is the absolute value of the Z-coordinate of the WALP-16 center of mass with origin placed at the bilayer center. The abscissa is the number of α-helical hydrogen bonds determined with a 3.5A cutoff on the hydrogen-oxygen distance and a 90 degree cutoff on the angle between the C-O and H-N vectors.

All points in the potential of mean force plane with counts at 7 or more temperatures were used to determine the thermodynamic formula:

$$\Delta G = \Delta E - T \Delta S$$

$$\Delta E = \Delta E_0 + \int_{T0}^{T} \Delta Cv dT'$$

$$\Delta S = \Delta S_0 + \int_{T0}^{T} \frac{\Delta Cv}{T'} dT'$$

$$\Delta Cv = \Delta Cv_0 + (d\Delta Cv / dT)_0 (T - T_0)$$

where the four parameters are: relative energy $\Delta E_0$ at temperature $T_0$, relative entropy $\Delta S_0$, relative heat capacity $\Delta Cv_0$, and the change in relative heat capacity with temperature $(d\Delta Cv/dT)_0$. Contours of $\Delta G$ are shown in figure 1. Contours of $\Delta E$ and $-T\Delta S$ are shown in figure 2. Points sampled at fewer than 7 temperatures are not shown.

**Acknowledgements:**
We thank A. Ladokhin, R. W. Pastor and S.H. White for enlightening discussions. This work was supported by the US DOE by LDRD project. Computer facilities were provided by Los Alamos Institutional Computing. In particular we thank D. Dawson, H. Marshall, M. Vigil, and C. Wampler for their assistance in porting code and using the Q machine at Los Alamos.

# Figures:

Figure 1: A folding trajectory for the WALP-16 is shown projected onto a two-dimensional surface.  The surface shows, with color and contour lines, the relative free energy changes along the changes in hydrogen-bonding of the helix and the z-position of the center-of-mass of the peptide relative to the bilayer center.  Color changes occur at 1kcal/mol intervals; solid contour lines are drawn at 2kcal/mol intervals.  All folding trajectories followed a similar route with initial stabilization of Trp at the interface followed by insertion and then folding.

Figure 2: Relative changes in enthalpy and entropy show the trade-offs that occur with peptide binding and insertion during the folding process.  Colors change at 10kcal/mol intervals; solid contour lines are drawn at 20kcal/mol intervals.  In particular, note the gain in enthalpic energy due to initial binding and the 7-11 kcal/mol gain in enthalpy due to hydrogen bond formation during the $\alpha$-helical folding within the bilayer interior.

Figure 3: The prevailing conceptual model of helical peptide insertion postulates that all transmembrane domains will fold within the interfacial zone and then insert as a folded domain into the bilayer.  The simulation results suggest, at least for this peptide, that the alternative pathway of partial binding at the interface, insertion to the bilayer interior and then folding to an $\alpha$-helix can be the preferred route for folding.

Figure 4: The electrostatic interaction energy between WALP and water as a function of WALP center of mass distance from the bilayer central plane (BLACK). Also shown is the number of bound waters versus position (BLUE) determined by counting all waters with an oxygen atom less than 4.0A in distance from any peptide atom. The bilayer central plane is positioned at z = 0 A, and the water-lipid interface is near z = 20 A. Even when the peptide is inserted deep into the membrane, it retains a significant interaction with the aqueous solvent, mostly through the existence of bound waters; consequently, it is not correct to think of the peptide as being completely desolvated even when it is in the center of the bilayer. These two figures were produced by combining the data at many different temperatures. This data was combined by first sorting the electrostatic interaction energy by temperature. At each temperature, the electrostatic energy versus position data was binned into overlapping bins of 5A width. The average electrostatic energy in each bin was assumed to have a linear variation with temperature. A least squares fit of energy versus temperature was used to determine the best estimate for the average electrostatic energy in each bin at 450K. Any temperature with fewer than ten sampled energies in a bin was not included in the linear fit. The same procedure was followed to compute the number of bound waters versus position. The vertical bars show estimated maximum errors.

Figure 5: The Coulomb interaction energy between the TRP residues and the bilayer as a function of their center of mass positions. The bilayer central plane is positioned at z=0, and the water-lipid interface is near z = 20 A.  The TRP residues have a strong electrostatic interaction with the lipid phosphatidylcholine groups that works to stabilize conformations with TRP in this region.  The electrostatic interaction energy between TRP and the membrane showed only a small temperature dependence, so all temperature data was combined and binned using a 2A bin width.  The vertical bar is an estimated maximum error.

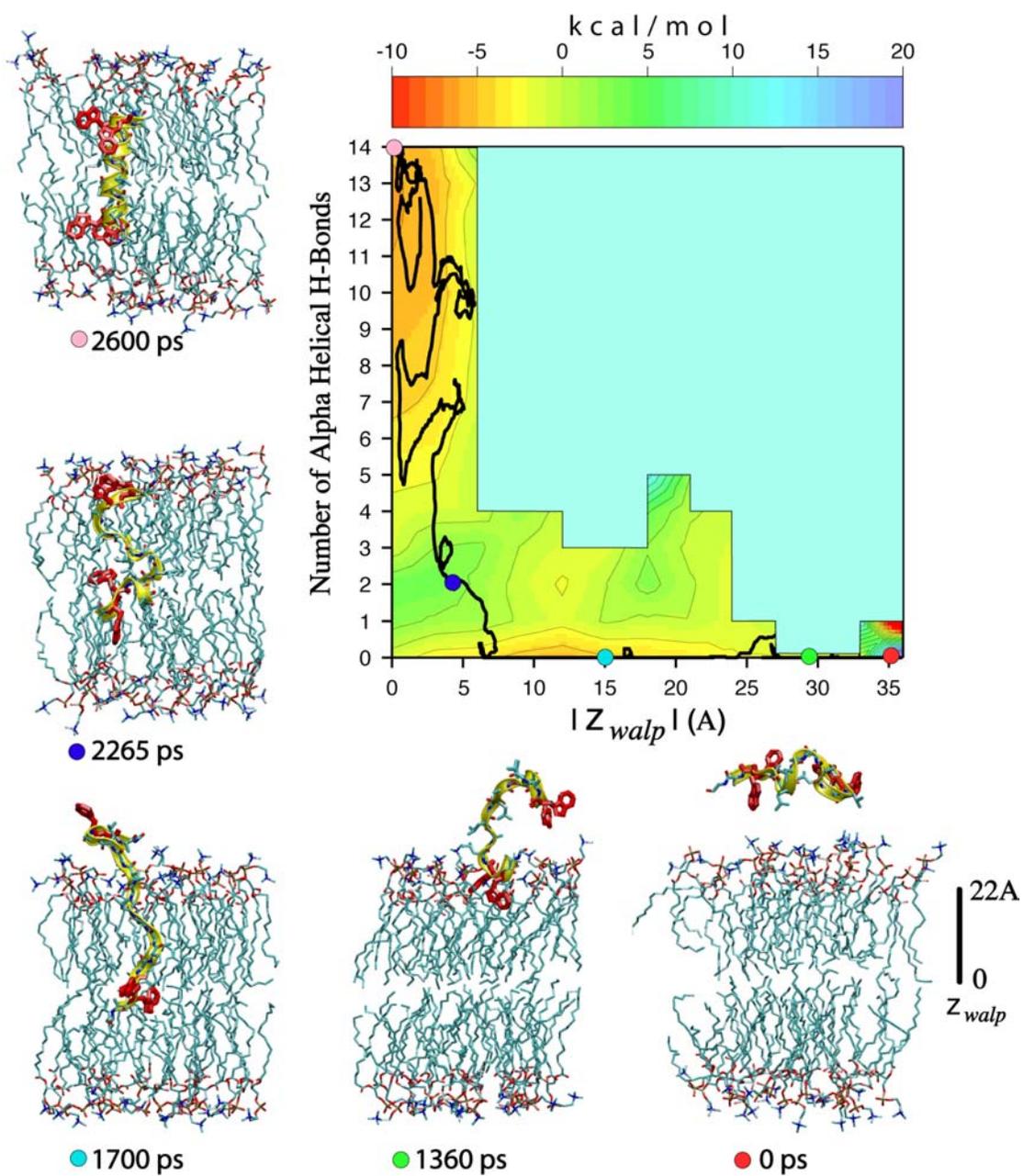

Figure 1.

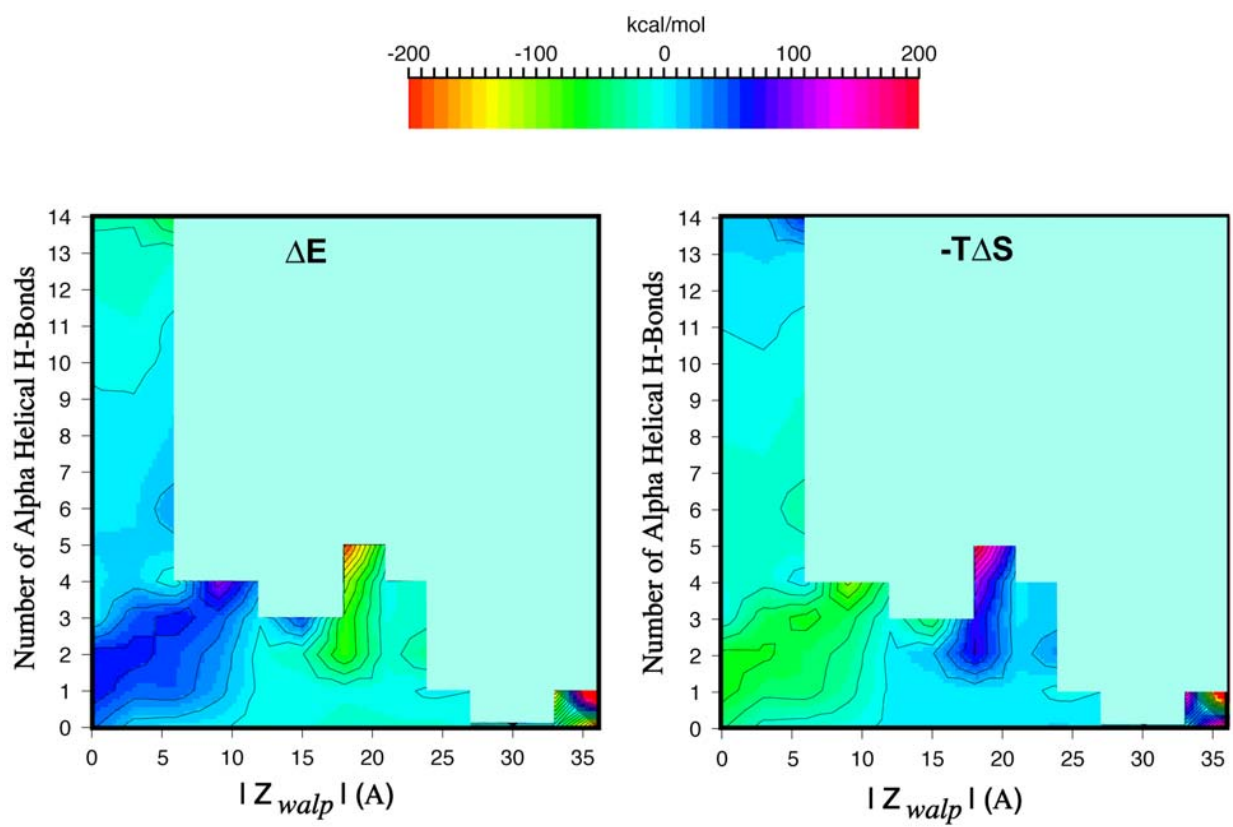

Figure 2.

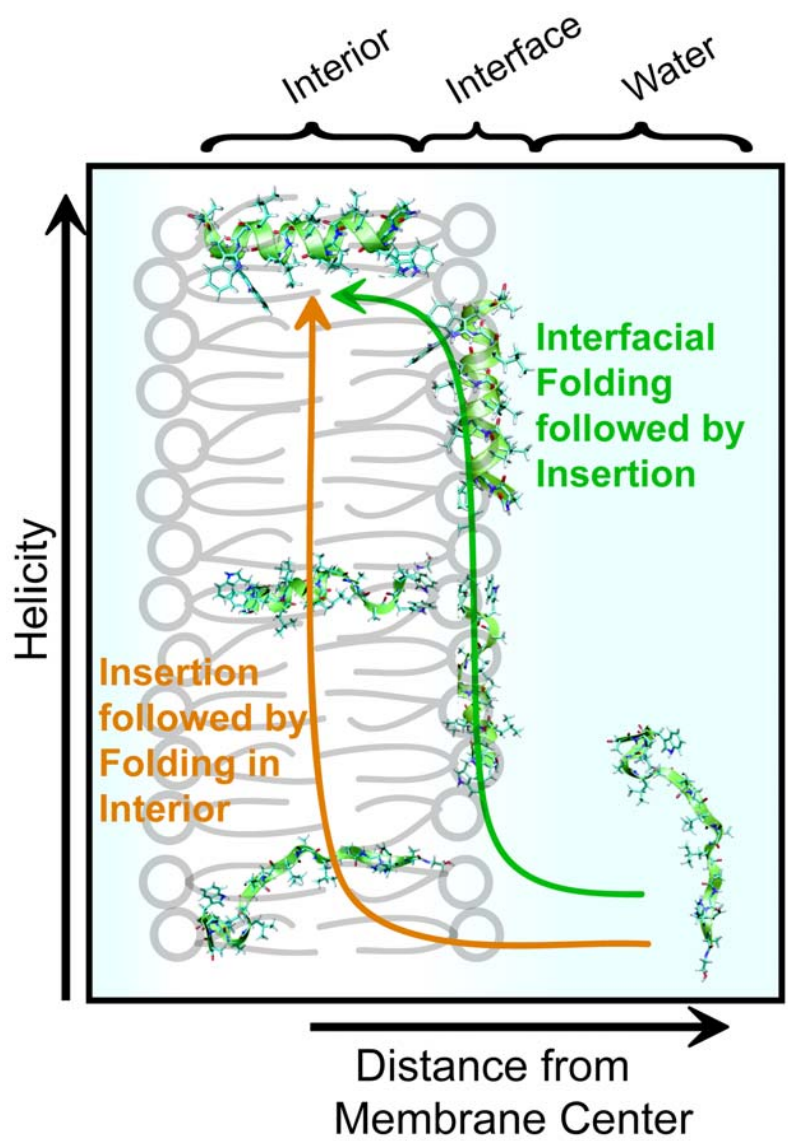

Figure 3.

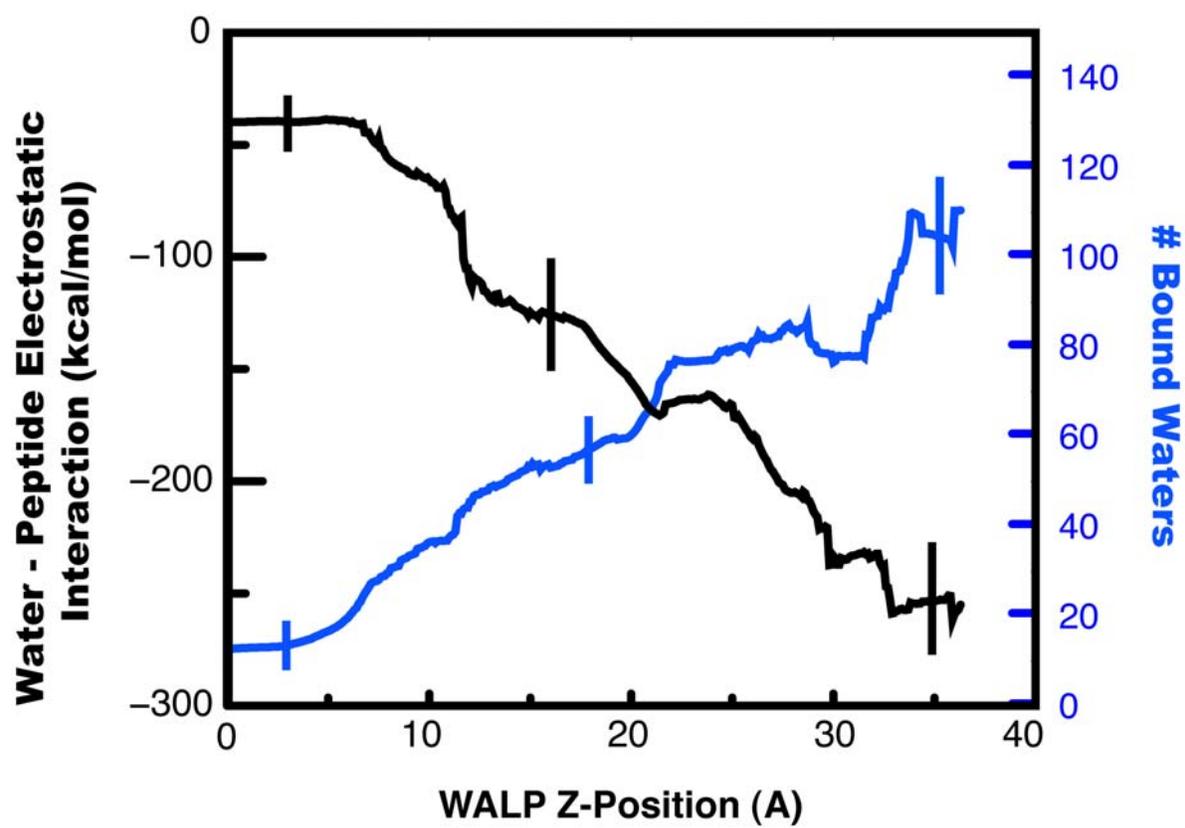

Figure 4.

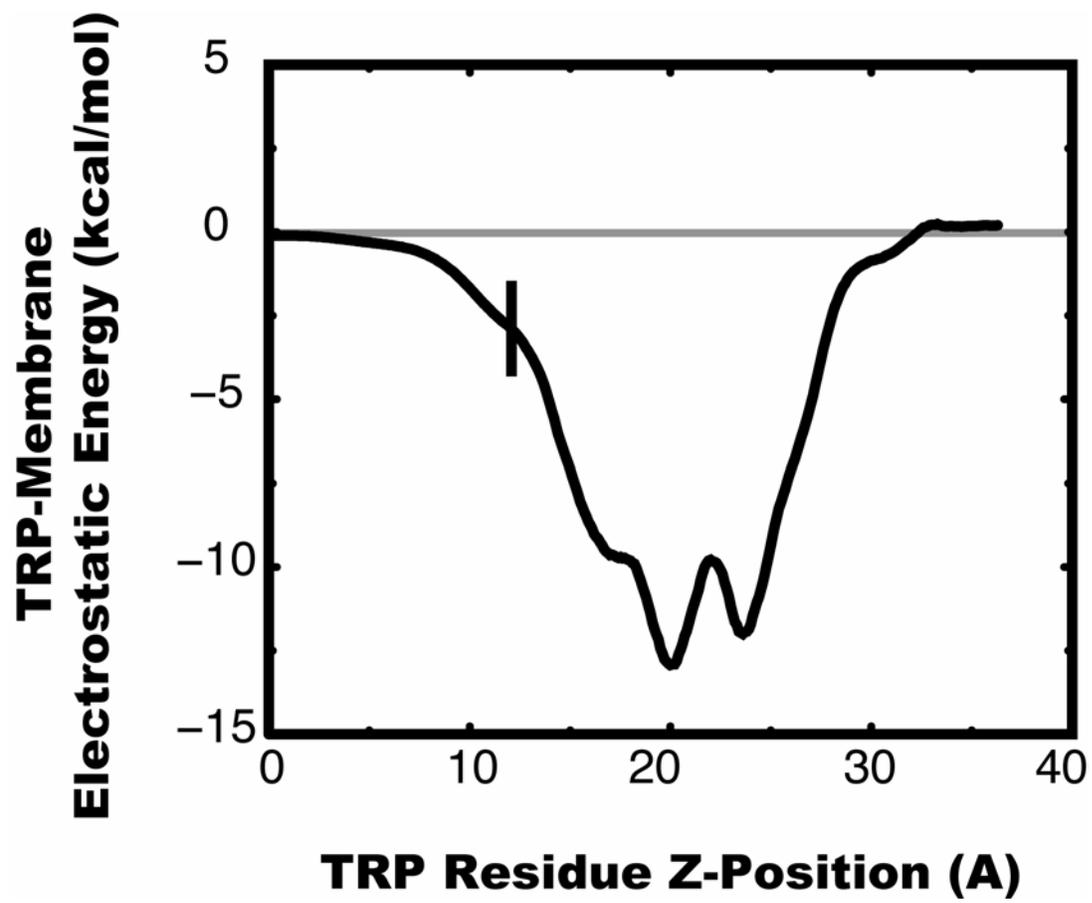

Figure 5.